# TMUML: A Singular TM Model with UML Use Cases and Classes

Sabah Al-Fedaghi

*salfedaghi@yahoo.com*

Computer Engineering Department, Kuwait University, Kuwait

## Summary

In the systems and software modeling field, a conceptual model involves modeling with concepts to support development and design. An example of a conceptual model is a description developed using the Unified Modeling Language (UML). UML uses a model multiplicity formulation approach, wherein a number of models are used to represent alternative views. By contrast, a model singularity approach uses only a single integrated model. Each of these styles of modeling has its strengths and weaknesses. This paper introduces a partial solution to the issue of multiplicity vs. singularity in modeling by adopting UML use cases and class models into the conceptual thinging machine (TM) model. To apply use cases, we adopt the observation that a use-case diagram is a description that shows the internal structure of the part of the system represented by the use case in addition to being useful to people outside of the system. Additionally, the UML class diagram is recast in TM representation. Accordingly, we develop a TMUML model that embraces the TM specification of the UML class diagram and the internal structure extracted from the UML use case. TMUML modeling introduces some of the advantages that have made UML a popular modeling language to TM modeling. At the same time, this approach supplies UML with partial model singularity. The paper details experimentation with TMUML using examples from the literature. Our results indicate that mixing UML with other models could be a viable approach.

## 1. Introduction

In scientific contexts, models are of fundamental significance. Consequently, a number of different types of models have been developed in the fields of software and system engineering. In systems and software modeling, a conceptual model involves modeling with concepts [1] used to support development and design. The model acts as an abstract framework of a phenomenon and of what is happening within the phenomenon that is to be represented.

1.1 Modeling Approach

In this paper, we adopt the notions that (a) conceptual models perform representational functions as symbolic depictions of a selected part of the world [2], and (b) Craik's [3] hypothesis that people think by manipulating models (internal representations) of the world. These ideas imply that models imitate people's internal thought processes, which in turn parallel reality. Mental models (or knowledge representations) are assumed to have the same essential features of symbolism (Craik's [3] term) that machines have. People translate external situations into static mental models and external events into mental events. Thus, external behaviors are mapped to mental behaviors. The terms "static," "events," and "behavior" will be defined later as elements of our conceptual thinging machine (TM) model. Fig. 1 clarifies this view, which involves external reality, a mental model, and TM modeling.

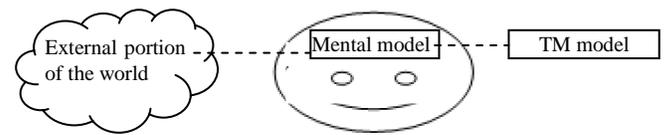

Fig. 1 Difference between mental and conceptual models.

In software and systems engineering, conceptual models (e.g., UML or TM) imitate mental models in structure and events. Typically, a concept is defined as a thing that is conceived by the mind [4]. We here want to emphasize that TM modeling also involves processing (*concepting*). Later, concept and concepting will be unified under one notion called a *thimac* (i.e., a thing/machine). Craik's [3] mental modeling can be put in the form of mental *thimacking*. That is, we suggest that thimacking is a mechanism for both mental and conceptual models.

Craik [3] proposed that manipulation of mental models of (portions) of the world that consist of (to use our terminology):

1. Mental modeling of some external process into an internal representation in thimac terms.
2. Deriving other mental models by some sort of inferential process.
3. Mentally modeling external events in terms of *thimacal* (thimacs plus time) events.

A TM model and its corresponding mental model represent the same underlying reality. TM has a relationship with reality through that mental model. For example, a TM model of a student registration system produces events similar to those that might have occurred in an actual (i.e., manual) process. In this sense, a TM model replaces its target system (e.g., a physical registration system). Because our interest is in software and systems engineering, we focus here on the system design models UML and TM, ignoring the issue of forming mental models.

The representational modeling style is an important aspect of system design because one can represent the same subject matter in different ways. According to Teller [5], "Nature may comprise no ultimate refinement of structure… the messiness of initial conditions, theoretical virtues such as simplicity, and other such constraints on theorizing are what really matters." The same phenomenon admits various formulations; hence, one solution is to characterize the phenomenon in terms of a set of models taken to apply to the world and parts of the world [5].

### 1.2 Research Problem

UML is an example of model multiplicity, wherein a number of models are used to represent alternative views. According to Lin et al. [6], the model multiplicity approach utilizes a distinct model for each view. Comprehending a system requires concurrent references to the various models and the creation of abstract associations that link them. Rather than being built into an integrating method that contains the various models, the alternative is to place such integration on the shoulder of the developers.

UML multiplicity [7] is a so-called quantitative multiplicity in which all objects are instances of a class. Only spatiality is counted; thus, the objects Mary, John, and Alice gain their multiplicity through their spatial *differences* (e.g., location and bodily features). The objects are all *united* as a class (i.e., humans). Singularity of objects is based on space (a common feature). This singularity of multiplicity based on spatiality works well because of the class/subclass mechanism. For example, cats and dogs are unified as subclasses of animals, thus achieving consistency. The multiplicities of cats and dogs are unified by the common features of animals. Of course, it is possible to have several independent super-classes that exist beside each other as long as their sets of objects do not intersect.

However, UML also has 14 types of modeling approaches—examples include special classes, use cases, sequence models, and activity diagrams. The multiplicity of objects (instances) of any of these models also needs an anchor of unity, similar to a class role for subclasses. All UML diagrams exist in two-dimensional spaces including the so-called behavior models (e.g., state diagrams). Because the instances of these models intersect with each other, the question becomes how to unify their multiplicity. Suppose a certain activity model and sequence model are applied repeatedly, producing a multiplicity of instances for each of them. What is the super-model for activity and sequence models to guarantee consistency? UML has none, but some models can be used as bridges between others.

There is another type of multiplicity that defines *separateness* in terms of time. In this case, objects are not unified based on spatiality but on temporality. An object is an instance of a class identified by its time of creation. The same class/subclass relationship is applied to achieve consistency between objects and subobjects. As we will see, TM modeling handles this time-related separateness by shifting instances to different levels of description. The basic claim here is that the notion of dynamic behavior is different in kind from the notion of static structure. Thus, a static structural description cannot be mixed with dynamic behavioral notions (events) in the same diagram, as is the practice in UML diagrams. In TM modeling, events are presented on a higher level than the static descriptions. In TM, a single model (singularity) is constructed first. Then it is partitioned into states, and an instance is repeatedly generated (multiplicity) by activating one partition at a time.

### 1.3 Proposed Approach

Software engineering has adopted the notion that multiple models are needed to represent and understand an entire system [8]. UML is a venture in this direction, as its model multiplicity embraces a family of design notations. In this paper, we attempt to bridge two perspectives on modeling by accepting some popular UML models, namely use case and class diagrams, as the bases for constructing a singular TM model. The resulting TMUML model includes use cases, class diagrams, and TM modeling.

In TM modeling, a use case is used along with a class diagram as an initial specification to build the TM model, as shown in Fig. 2. Thus, we aim to incorporate use cases as the basic blocks for system specification and provide a foundation by facilitating the elicitation, collection, analysis, and documentation of requirements [9] [10]. Additionally, the class diagram is used to model the structural view of a system that includes the abilities to carry data and execute actions.

### 1.2 Outlines

The next section is a brief review of TM modeling (See [11] and its TM references by the author). Section 3 focuses on use case modeling. Section 4 gives a TMUML case study.

## 2. Thinging Machine Modeling

The TM world is a world of *thimacs*: things that are simultaneously machines. Bryant [12] says of such a thesis, "In short, being is an ensemble or assemblage of machines." All things are created, processed, and transported (acted on), and all machines create, process, and transport other things (act; Fig. 2). Beings in the world have two roles: they can serve as machines, which act as subjects, and things, which act as objects. This is what we understand from Aristotle: "'being' is in one way divided into individual things and is in another way distinguished in respect of potency" [13].

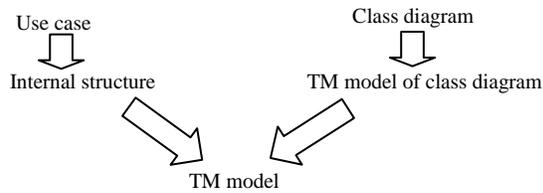

Fig. 2 Overview of our approach.

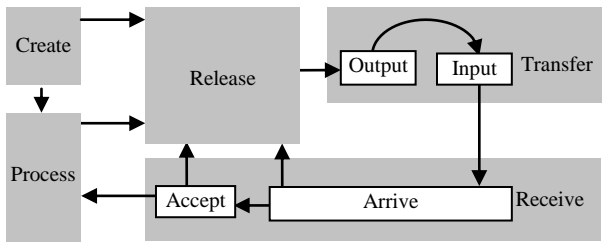

Fig. 3. A thinging machine.

Thimacs have a dual nature: They are atemporal and temporal (the classical duality of thing vs. object). The machine side of a thing includes spatiality and actionality (generic actions) that embed the potentiality for change. The machine side of an object is shown in Fig. 3. TM can be described as the following generic (basic) actions:

**Arrive:**   A thing moves to a machine.
**Accept:**   A thing enters the machine. For simplification, we assume that all arriving things are accepted; hence, we can combine the *arrive* and *accept* stages into one stage: the **receive** stage.
**Release:**  A thing is ready for transfer outside the machine.
**Process:**  A thing is changed, but no new thing results.
**Create:**   A new thing is born in the machine.
**Transfer:** A thing is input into or output from a machine.

Additionally, the TM model includes memory organization, which plays the role of storage for each action. For simplification purposes, one may assume each TM has a single storage area. Additionally, the TM model includes the mechanism of triggering (denoted by a dashed arrow in this study's figures), which initiates a flow from one machine to another. Multiple machines can interact with each other through movement of things or triggering. Triggering is a transformation from one series of movements to another.

## 3. Use Cases

Because we have elaborated on class diagrams in a previous paper [11], where class diagrams are formulated in terms of the TM model, this section focuses on use cases and their incorporation in TM modeling.

### 3.1 About Use Cases

Use cases are useful to people outside of the system. They are used to specify system behavior from the user's point of view. Applications are conceived in terms of use cases that explain what the stakeholder expects from the system by means of describing an interaction with it [14]. Additionally, use cases serve as basic blocks for system specification in software and systems engineering processes and facilitate the elicitation, collection, analysis, and documentation of requirements [9] [10].

Use-case-driven analysis is called the "cornerstone" of software and systems modeling in UML and SysML. In software engineering, no other construct as significant as use cases has been adopted so quickly or so widely among practitioners because use cases play a role in so many different aspects of software engineering [15].

In UML 2, "a use case is the specification of a set of actions performed by a system, which yields an observable result that is, typically, of value for one or more actors or other stakeholders of the system" [16]. A use case may be specified by means of a full description of interactions using an elaborated textual form with low-level pseudo-code resulting in well-known problems. An improvement to this approach is the use of diagrammatic forms to specify permitted interactions [14].

According to Jacobson [15], the application of use cases is not limited to software development. They can help to understand business requirements, analyze existing business, design new and better business processes, or exploit the power of IT to transform business. By using use cases, we can identify the ways in which the systems will affect a business and which systems are needed to support it. Use cases are not a uniquely object-oriented defining characteristic, although they could be used with practically any software development approach [17]. Many people think use cases are only applicable to user-intensive systems, but the original idea for use cases came from telecom switching systems, which have both human and machine users. However, use cases are applicable to all systems that are used [15].

### 3.2 Use Case as a Sketch of the Modeled System's Internal Structure

One objective of conceptual modeling is identifying the right problem (we might refer to this as "problem structuring") as well as understanding the system and its boundaries [18]. From a designer's point of view, as suggested by da Silva [19], a use-case diagram shows the internal structure and functional decomposition of a model [10]. Specifically, according to Isoda [10], "actors and a subject that appear in a use-case diagram are also part of the behavior description because they appear in it… a use-case diagram represents the internal structure of a use case's behavior description." To illustrate the meaning of the internal structure that can be extracted from a use case, Fig. 4 shows an example of a use case given by Dijkman and Joosten [20]. Fig. 5 shows different regions of the TM model that correspond to the use case. The solid arrow denotes communication interactions, as in the use case given. The dashed arrows denote triggering in the TM sense.

Fig. 6 shows another example of a use case with a more complex internal structure, with <include>, <extend>, and generalization. In the internal structure shown in Fig. 7, each actor is a region: the customer is represented by the space on the left and the banking application system is represented by the space on the right (Circle 1). The login system is structured as one region (blue box), containing login (2), which triggers either verify password (3) or continues in a sequential manner. The red subregion shows the sequential program (5): verify password (3) may trigger error (4) or be followed by receiving a transaction from the customer to trigger check balance (6), transfer funds (7), or make payment (8). Verify sufficient funds (9) is performed when checking the balance and transferring funds. Make payment (10) involves another interaction with the customer to determine the type of account (11 and 12).

Fig. 7 is extracted from the use case except when introducing the more generalized sub-actor transaction. This was implied by the use case sharing a group of functions between the customer and bank, as shown in Fig. 8.

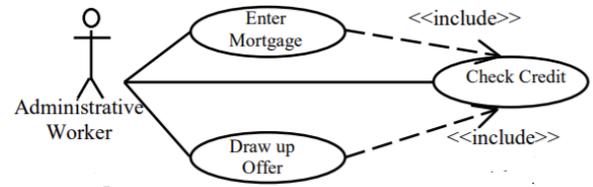

Fig. 4  Partial sample of a use case (from Dijkman and Joosten [28]).

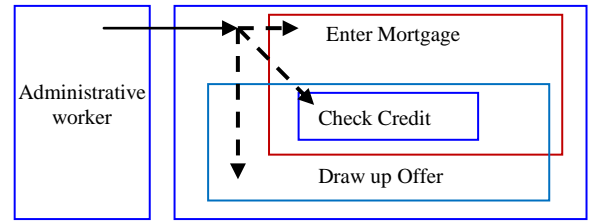

Fig. 5 Illustration of the internal structure conveyed by a use case.

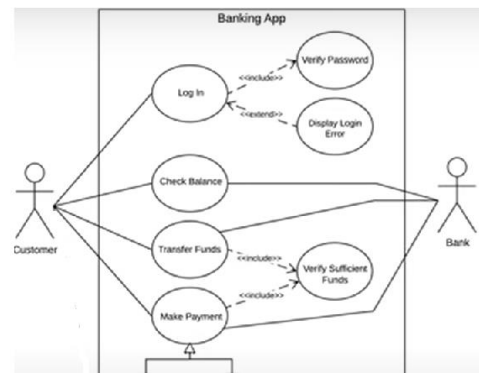

Fig. 6  A partial use case (from www.programmersought.com).

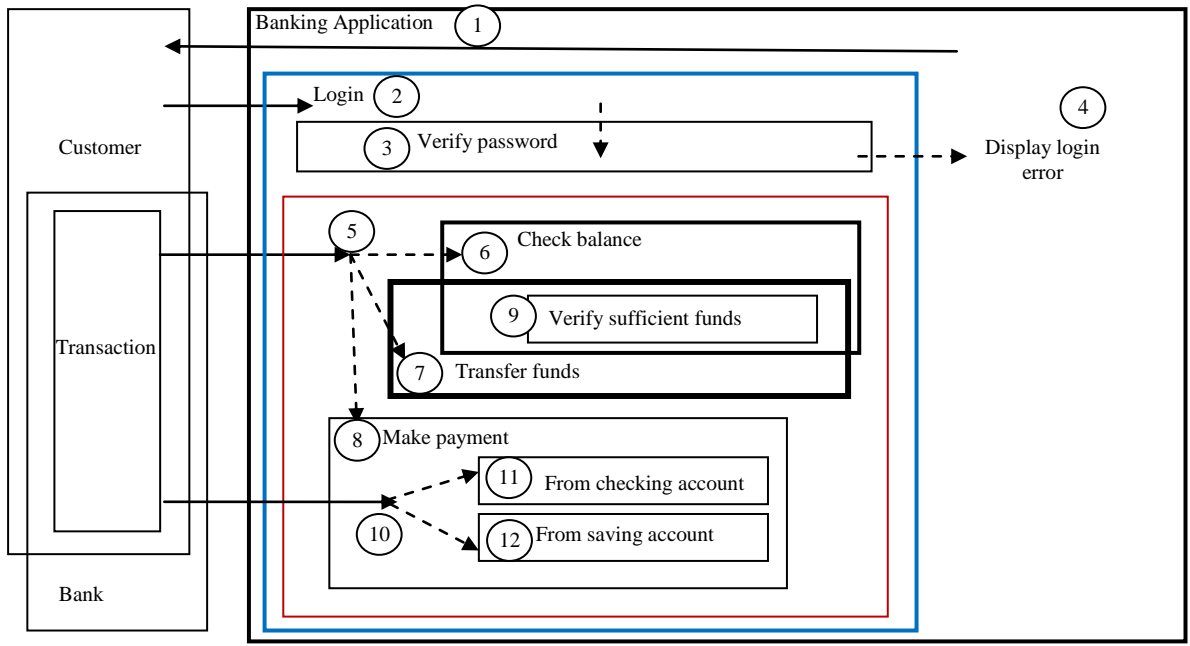

Fig. 7  Internal structure inferred from the banking application use case.

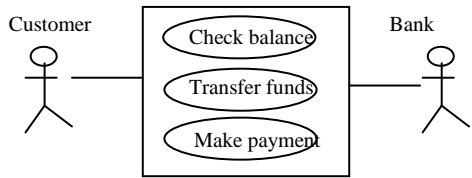

Fig. 8  Alternative representation of actors sharing use-case functions.

## 4. Building a TMUML Model

According to da Silva [19], specifying system requirements facilitates communication among stakeholders and supports development processes. Generating such specifications requires a systematic, rigorous, and consistent method with a large set of constructs at different levels of detail and different types of requirements (e.g., goals, functional requirements, constraints, and use cases). This "large set of constructs" in [19] seems to refer to UML-style notations. According to [19],

- A modeling language such as UML (and SysML at some extent) is a reasonable foundation for requirements modeling, but it is incomplete for modeling requirements because it lacks models that tie requirements to business value and models that present the system from an end user's point of view.
- A way to establish relationships between constructs defined for different types of diagrams has not been well defined (e.g., between use cases and classes or between state machines and use cases).
- UML does not have a standard way of further classifying its constructs (e.g., neither use cases nor actors have any further semantics).

Next, da Silva [19] proposed a UML-like requirements specification language that solely focuses on constructs that "most related with use cases approaches, i.e., focused on the following constructs: use cases, actors, data entities, state machines, and their inherent relationships." Therefore, [19]'s UML-like project is actually similar to our venture in this paper, where UML is adopted partially. We here opt to compare our approach to such a proposal, and UML in general, by modeling a problem side-by-side in UML and TM modeling and letting the reader contrast the advantages and disadvantages of both methodologies.

According to Aguirre-Urreta and Marakas [21], papers comparing diagrammatic conceptual models (e.g., entity relationship and UML/object-oriented modeling techniques) in the published literature, although "vibrant," have "often yielded *equivocal* findings." In this context, Houy et al. [22] claim that model understandability remains "ambiguous [and] research results on model understandability are hardly comparable and partly imprecise." One way to contrast conceptual models can be through experimentation. For example, Valaski et al. [23]

used eight professionals and 80 students to evaluate the expressiveness of UML and OntoUML.

The point here is that it is very difficult to present a detailed comparison between UML and TM modeling, especially because the latter is still a mere proposed approach. Achieving a reasonable level of comparability at this stage of development involves modeling the same problem in UML and TM and contrasting the diagrammatic representations side-by-side in a way that can be grasped by non-technical persons. After all, the two approaches may be compatible, a topic that is explored in this paper.

Following this strategy, Fig. 9 illustrates an example of a use case of an invoice management system expressed in TM notations. The purpose of the figure is to contrast the UML use cases' diagrammatic form with their TM diagrammatic representation, not to provide a fair review.

The question is how to represent this use case in TM modeling. As we mentioned previously, the use case helps in identifying the internal structure of the system. Accordingly, the actors are the manager, operator, and customer. The system is called Manageinvoices (Fig. 10), as we use the same names given in the use case. Accordingly, Fig. 10 shows the internal structure discussed in the previous section in preparation to construct the TM representation.

In Fig. 10,

- The operator (Circle 1) interacts with the system to trigger (activate; Circle 2) six machines (3-8). This interaction involves inputting data and activating (triggering; dashed arrow) some methods.

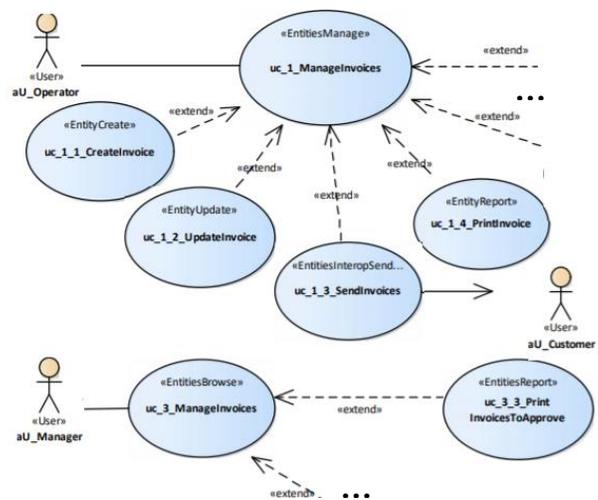

Fig. 9  Use case of invoice management system (partial from [19]).

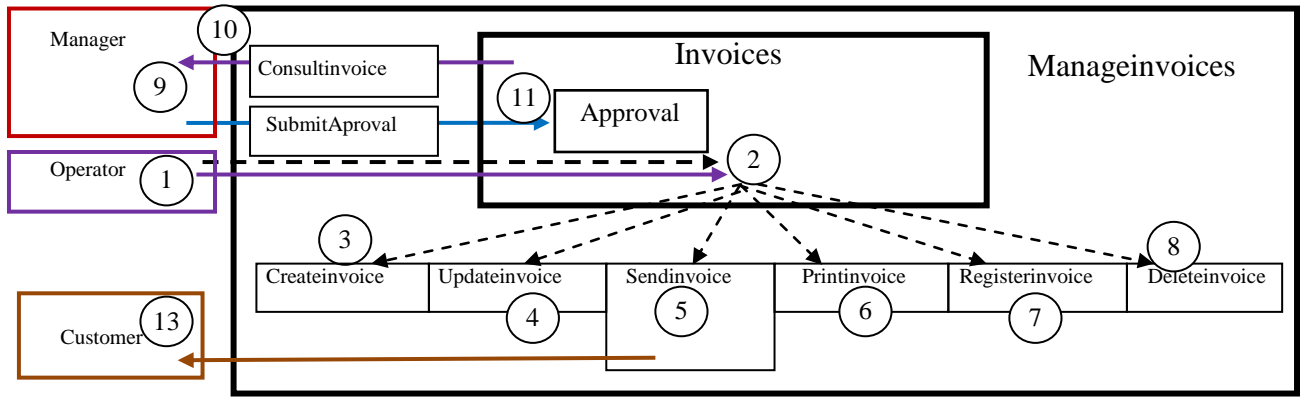

Fig. 10 The internal structure of the TM model is provided by the given use case.

- The manager (9) interacts when receiving the new invoice (10) by approving/disapproving the invoice (11).
- The customer (13) interacts by activating the *sendinvoice* machine (5), which sends a copy of the invoice to the customer.

This internal structure of processing an invoice gives a rough basis for actors and their interaction with the invoice management system.

We refine Fig. 10 as shown in Fig. 11, which sketches different flows and types of triggering to and from the invoice class (machine; Circle 1). First, the actions of the operator are listed (2), which result in a flow of data (an attribute value) or triggering an operation (e.g., delete). The operator or customer may request outputting the invoice (3) (e.g., to a printer). The manager who receives the newly created invoice sends back approval/disapproval as an attribute value in the invoice (4 and 5).

Additionally, da Silva [19] gives the class diagram of the system as shown partially in Fig. 12. Fig. 13 shows the corresponding TM diagram, in which we make the following assumptions to save space:

- Only the class invoice (Circle 1 in Fig. 13) is modeled, since all classes can be represented in the same way (e.g., customer class [2]).
- In the invoice class, only the invoice attribute ID (3) is modeled, since all other invoice attributes (4) are represented in a similar way.

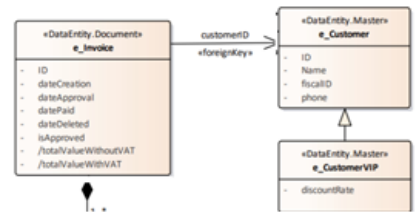

Fig. 12 Class diagram of the given example (partial from [19]).

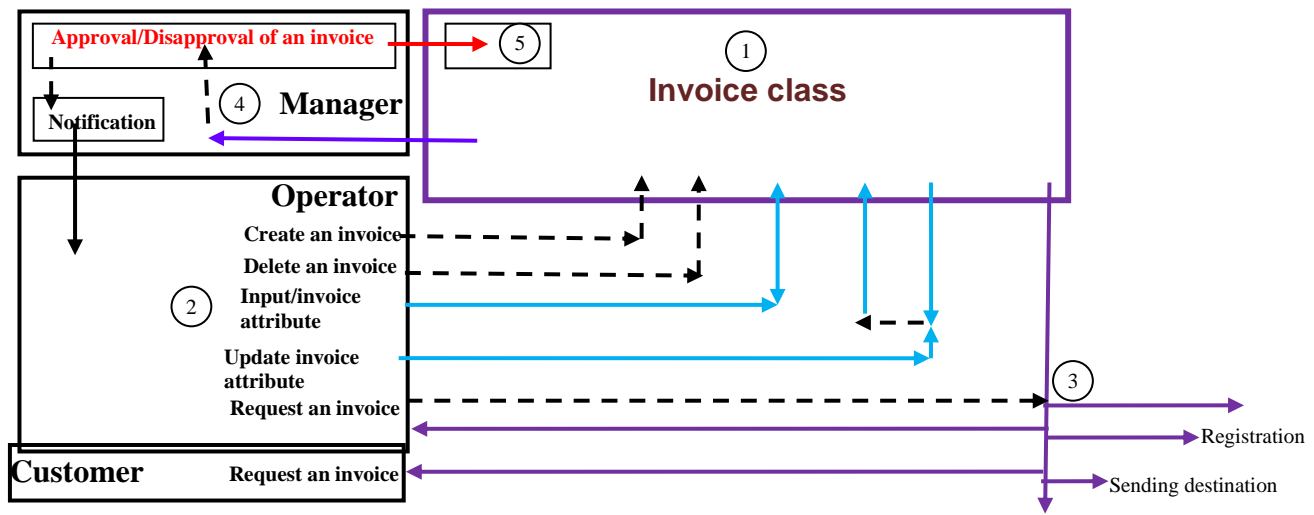

Fig. 11 Sketching different flows and types of triggering to/from the invoice.

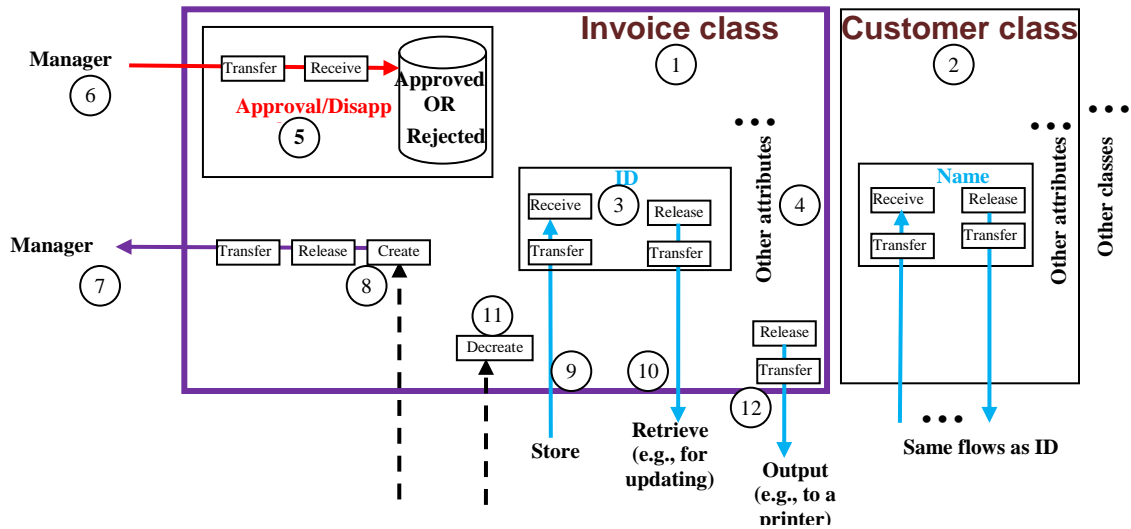

Fig. 13 TM representation of the classes.

- The approval/disapproval attribute (5) is unique in the invoice diagram. Its value is set by the manger (6) who receives the new invoice (7) after creating it (8).
- The value of the ID can be stored (9) and retrieved (10).
- The instance of the invoice can be deleted (decreated, the inverse of create; 11) or downloaded (12) to be sent somewhere, such as a printer.

## 4.1 Static Model

Fig. 14 shows the static TM model that unifies the use case and class diagrams of the invoice management system into a single diagram. The figure shows the static model of the inventory management system with the following aspects:

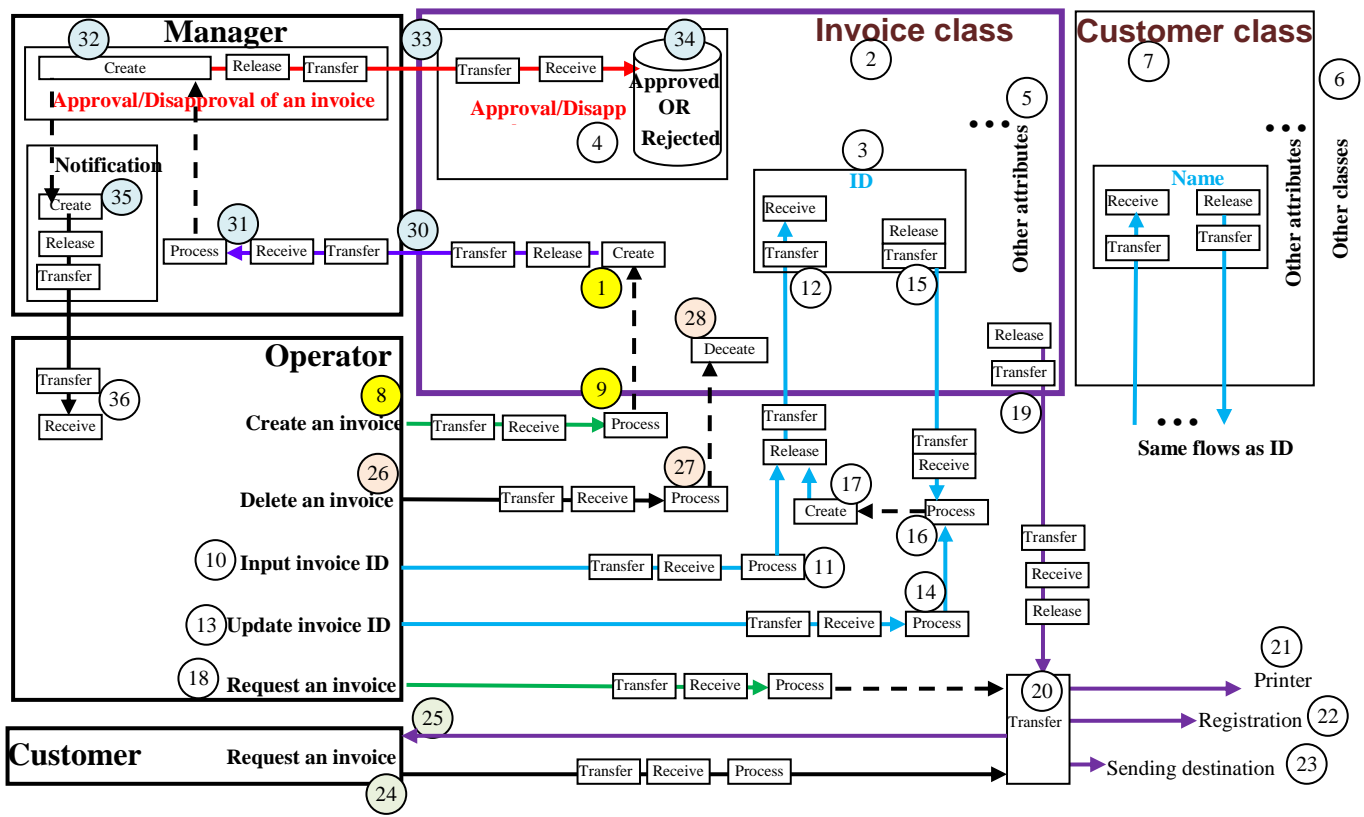

Fig. 14 TM model of the invoice management system.

- The figure features *classes*, which allow the creation (Circle 1) of the invoice class (2) with two attributes: ID (3) and approval/disapproval (4). Other attributes can be added (5) and treated in a similar way as ID. Other classes (6) – for example, the customer class (7) – can also be modeled in a way similar to the invoice class. The relationships among classes are not introduced in this paper.

- *Creating an instance of invoice* starts with a request from the operator (8) that is received and processed by the system (9) to trigger *create* in the invoice structure (1). This *create* action may trigger inputting values for attributes, as in an updated invoice ID that will be specified next. Or, *create* might put nulls as initial values for a class instance, as when creating an object in object-oriented languages.

- *Input invoice ID* (10) facilitates the operator inputting values for attributes (11 and 12). It can be used for all attributes of the invoice to realize the creation of an invoice. It can also be realized at the operator interface level by filling in all values at once (e.g., filling the interface page). The details of how to fill the values of an invoice can be added to the TM model.

- *Update invoice ID* (13) can be used to change a current attribute value that is received (14). We assume that this process requires retrieving the current value of an attribute (15), comparing the two values (16), creating a new value (17), and depositing this new value in the attribute (12).

- *Request an invoice* (18) provides the operator with an instance of an invoice. This process requires the retrieval of an invoice instance (19) from the system and transfers it (20) to its destination (21, 22, and 23).

- *Request a copy of the invoice* (24 and 25) in the customer region is similar to the previous item.

- *Delete an invoice* (26 and 27) is similar to the *create an invoice* process. It is a reversed version of creation, so it is labeled *decrease* (28).

- *Approval/disapproval of the invoice* (after being created) requires sending (30) the invoice instance to be examined (31) by the manager, who then inputs the value approved or disapproved into the invoice instance (32, 33 and 34). Additionally, the manager sends a notification (35) about his or her decision to the operator (36).

Note that the static model of Fig. 14 follows the internal structure of the use case structure (Fig. 10). In this paper, the modeling process involves moving from a traditional use case diagram to extract the internal structure of the system, then moving to modeling using TM notations. Therefore, practically speaking, the use case and class diagrams are used as a first step to develop the TM diagram (in agreement with Fig. 2 in Section 2). Additionally, note that the attributes and methods of the class in Fig. 14 are spread throughout the TM diagram. However, the methods are integrated in the TM diagram at a higher level of description (i.e., over Fig. 14), as will be shown next in the dynamic model.

## 4.2 The Events and Behavior Model

An event in a TM can be represented by its region. Accordingly, the static model of the invoice management system can be partitioned into regions of events $E_1$, …, $E_{21}$ as shown in Fig. 15.

$E_1$: The operator requests the creation of a new invoice (instance) that is received and processed by the system.

$E_2$: The system creates an invoice.

$E_3$: The operator requests the deletion of an invoice (instance) that is received and processed by the system.

$E_4$: The system deletes an invoice (decrease is a version of create that reverses the create action).

$E_5$: The operator inputs attribute ID values that are received by the system.

$E_6$: The ID attribute value is stored.

$E_7$: Invoice ID is to be updated (e.g., error in the current value).

$E_8$: The current ID value is retrieved.

$E_9$: The input and current ID values are processed, and a new value is created.

$E_{10}$: The operator requests that an invoice be sent.

$E_{11}$: The invoice is retrieved.

$E_{12}$: The invoice is sent to a destination.

$E_{13}$: The invoice is sent for registration.

$E_{14}$: The invoice is sent to the printer.

$E_{15}$: The customer requests an invoice, and the system sends that invoice to the customer.

$E_{16}$: The invoice flows to the manager.

$E_{17}$: The manager approves or disapproves the invoice.

$E_{18}$: The approval or disapproval value flows to the system to be stored.

$E_{19}$: The manager sends an approval/disapproval notification to the operator.

For the sake of completeness, we add the following events:

E: The event of the system being active and beginning interaction with different users.

$E_O$: The event of the operator interaction session.

$E_C$: The event of the customer interaction session.

$E_M$: The event of the manager interaction session.

Accordingly, Fig. 16 shows the behavior model of the invoice management system. The methods specified in [19]'s class diagram (Fig. 12) can be defined as follows.

> Createinvoice: $E_1 \rightarrow E_2$
> Updateinvoice: $E_5 \rightarrow E_6$ (for all attributes of the class)
> Sendinvoice: $E_{10} \rightarrow E_{11} \rightarrow E_{12}$
> Printinvoice: $E_{10} \rightarrow E_{11} \rightarrow E_{13}$
> Registerinvoice: $E_{10} \rightarrow E_{11} \rightarrow E_{14}$
> Deleteinvoice: $E_3 \rightarrow E_4$

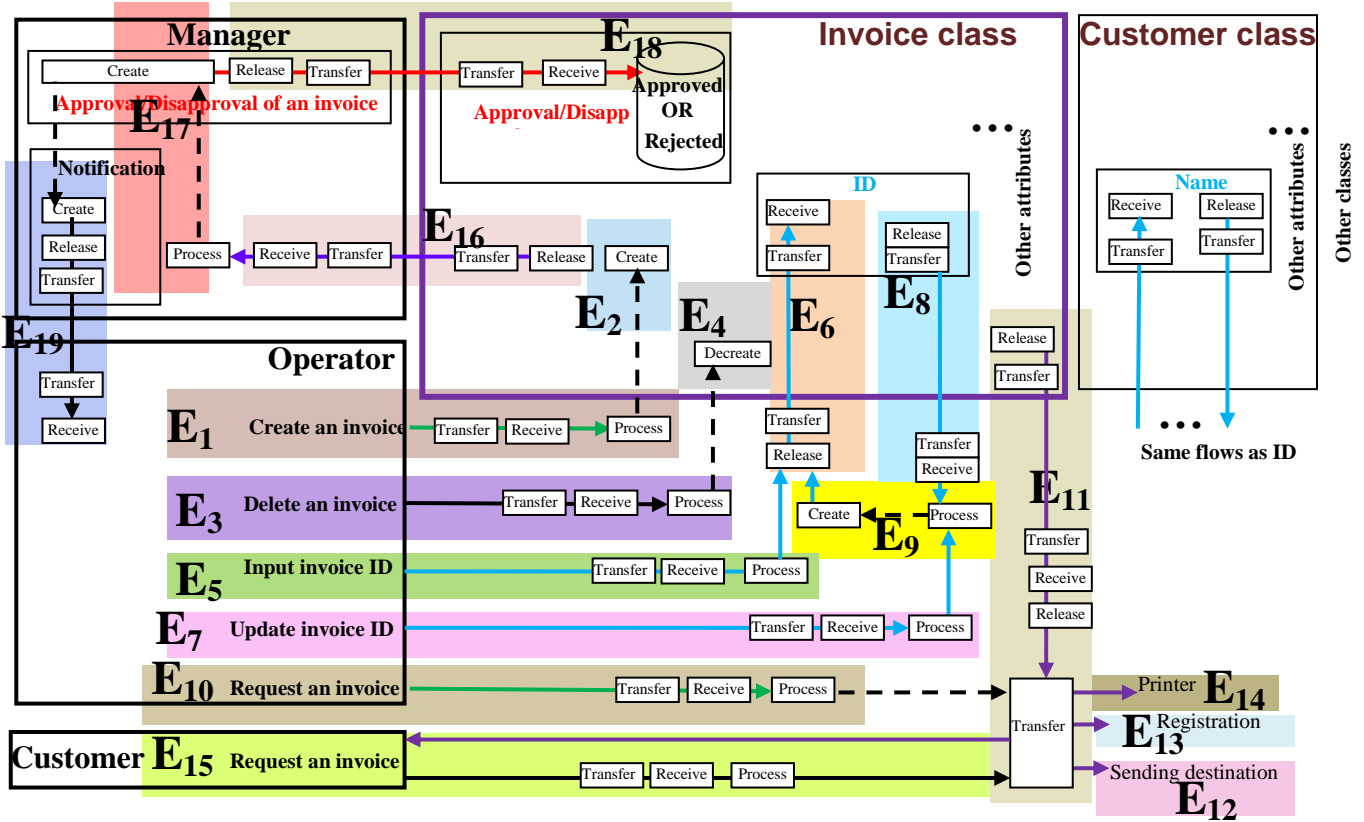

Fig. 15  Event model of the invoice management system.

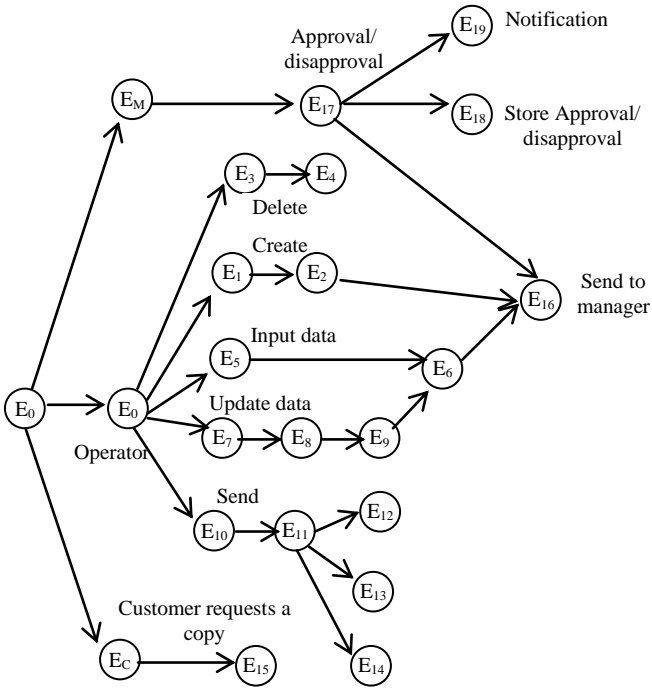

Fig. 16  Behavioral model of the invoice management system.

### 4.3 State Diagram

Last, da Silva [19] introduces a state diagram for the invoice system. Apparently, something is still missing in the model that must be supplemented by a state diagram to specify the invoice's behavior given in Fig. 17. Other classes/objects also need such behavior specifications. The figure shows the specification of the invoice's behavior through the definition of its respective state machine. Fig. 5 depicts the equivalent "UML-like" representation for the example. As described by [19], "This state machine includes six states: Initial, Pending, Approved, Rejected, Paid and Deleted. Initial is an initial state, while Paid and Deleted are final states." In TM modeling, such a description of behavior is not necessary because we already specified the behavior in Fig. 16.

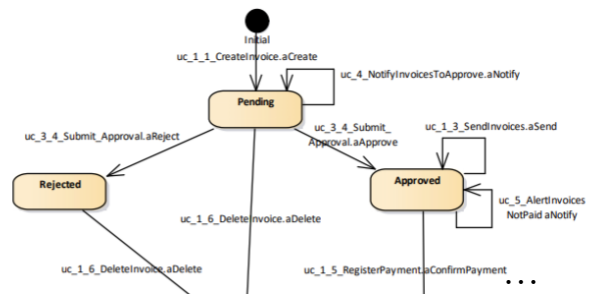

Fig. 17  State diagram (adapted from [19]).

## 8. Conclusion

This paper presents a possible solution to researchers who are still not satisfied with all UML apparatuses but who also view use cases as basic blocks for system specification. We provide a foundation by facilitating the elicitation, collection, analysis, and documentation of requirements. Furthermore, we consider the structurability of a UML class diagram a foundation for further software system development (e.g., design phase). We propose the singular TM model based on use cases and class diagrams (Fig. 18). This venture is an interesting experiment in mixing parts of UML with other models. This is an application of UML philosophy, because it is not necessary to utilize all models in UML. The result of applying the proposed approach seems promising. Further research would clarify related issues, such as class relations.

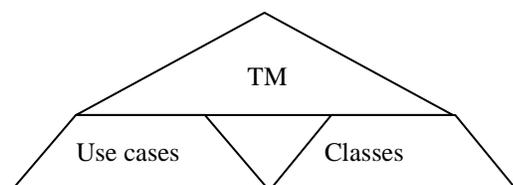

Fig. 18  General view of the proposed approach.